# Universal Scaling in Intrinsic Resistivity of Two-Dimensional Metal Borophene

Jin Zhang[1,4,†], Jia Zhang[1,4,†], Cai Cheng[1], Jian Liu[1,4], Johannes Lischner[2], Feliciano Giustino[3], and Sheng Meng[1,4,5,*]

[1]*Beijing National Laboratory for Condensed Matter Physics, and Institute of Physics, Chinese Academy of Sciences, Beijing 100190, P. R. China*

[2]*Departments of Materials and Physics, and the Thomas Young Centre for Theory and Simulation of Materials, Imperial College London, London SW7 2AZ, United Kingdom*

[3]*Department of Materials, University of Oxford, Parks Road OX1 3PH, Oxford, United Kingdom*

[4]*School of Physical Sciences, University of Chinese Academy of Sciences, Beijing 100049, P. R. China*

[5]*Collaborative Innovation Center of Quantum Matter, Beijing 100190, P. R. China*

[†] These authors contribute equally to this work.

* Correspondence: smeng@iphy.ac.cn

# Abstract

Two-dimensional boron sheets (borophenes) have been successfully synthesized in experiments and are expected to exhibit intriguing transport properties such as the emergence of superconductivity and Dirac Fermions. However, quantitative understanding of intrinsic electrical transport of borophene has not been achieved. Here, we report a comprehensive first-principles study on electron-phonon driven intrinsic electrical resistivity ($\rho$) of emerging borophene structures. We find that the resistivity is highly dependent on the atomic structures and electron density of borophene. Low-temperature resistivity of borophene $\rho$ exhibits a universal scaling behavior, which increases rapidly with temperature $T$ ($\rho \propto T^4$), while $\rho$ increases linearly for a large temperature window T > 100 K. It is observed that this universal behavior of intrinsic resistivity is well described by Bloch-Grünesisen model. Different from graphene and conventional three-dimensional metals, the intrinsic resistivity of borophenes can be easily tuned by adjusting carrier densities while the Bloch-Grünesisen temperature is nearly fixed at ~100 K. Our work suggests monolayer boron can serve as an intriguing platform for realizing high-tunable two-dimensional electronic devices.

# Introduction

The resistivity of metals originating from electron-phonon (*e-ph*) interactions (*i.e.* intrinsic resistivity or $\rho_{e\text{-}ph}$) is a fundamental quantity in condensed matter physics and materials science. At finite temperatures, scattering of electrons by phonons is generally the dominant source of resistivity. In a typical three-dimensional metal with a large Fermi surface, the "intrinsic resistivity" is proportional to the temperature (*T*) at high temperatures, attributed to the bosonic nature of the phonons.[1-3] Below a critical point, the resistivity is expected to decrease more rapidly with the drop of temperature, $\rho_{e\text{-}ph} \propto T^5$. In a two dimensional (2D) conductor, low-temperature $\rho_{e\text{-}ph}$ is proportional to $T^4$ due to the reduced dimensionality. The transition point between high-temperature and low-temperature regimes is determined by the Debye temperature ($\Theta_D$). The $\Theta_D$ is the typical temperature, upon which all phonons are excited to scatter carriers.[3] However, in systems with low-density electrons, the Fermi surface is much smaller than the size of the Brillouin Zone (BZ). It results in a different characteristic temperature in the systems with a small Fermi surfaces, the Bloch-Grüneisen temperature $\Theta_{BG} = 2\hbar v_s k_F / k_B$ (where $k_F$ indicates the size of the Fermi surface; $v_s$, $\hbar$ and $k_B$ are the sound velocity, reduced Plank constant and Boltzmann constant, respectively).[3-6] The temperature $\Theta_{BG}$ is much smaller than $\Theta_D$ in low-density electron gases, and the low-temperature behavior of intrinsic resistivity is summarized as the Bloch-Grüneisen model [1,2].

As a semimetal with the largest known electrical conductivity, graphene provides a textbook example for transport properties in 2D systems. The intrinsic electrical resistivity of graphene arising from electron-phonon interactions has been explored both experimentally and theoretically.[5-7] Low-temperature $\rho_{e\text{-}ph}$ of graphene varies linearly with $T^4$, while at high temperatures $\rho_{e\text{-}ph}$ is proportional to *T*. The transition point of these two distinct regimes is determined by the $\Theta_{BG}$ temperature as the result of its point-like Fermi surface[8-9]. Because of the dependence on $k_F$, $\Theta_{BG}$ of graphene is highly tunable by varying the carrier density or Fermi energy[5]. Previous experiments[5] by Efetov and Kim have confirmed the temperature $\Theta_{BG}$ can be tuned by almost an order of magnitude by applying a gate voltage. Park *et al.* have demonstrated the relative role of the phonon modes (acoustic or optical) and the microscopic nature of *e-ph* interactions in $\rho_{e\text{-}ph}$.[6] However, the absolute value of $\rho_{e\text{-}ph}$ on the order of 1.0 μΩ*cm at room temperature, is not sensitive to the applied external carrier densities, which may limit its potential applications in highly-tunable nano devices.

Different from graphene, 2D boron sheets, namely borophenes, have a variety of polymorphs, which can be regarded as a triangular lattice with periodic vacancies[10-16]. Recently, several borophene phases have been synthesized on Ag surfaces, *e.g.*, $\beta_{12}$, $\chi_3$ and triangle sheets [17-20]. All of the experiment-realized borophenes exhibit intrinsic metallic properties, providing an ideal platform to explore the transport properties of 2D metals, in addition to 2D semimetals (such as graphene) and 2D semiconductors (such as $MoS_2$). Experimental works have revealed the existence of Dirac cones in $\beta_{12}$ and $\chi_3$ sheets[21-22]. Moreover, theoretical works have demonstrated a variety of novel properties of borophenes, such as phonon-mediated superconductivity[16,23-25] with a critical temperature ~20 K and excellent mechanical behaviors *etc.*[26-33] As a 2D elemental metal, the intrinsic electrical resistivity of borophene lies at the heart of the potential application in electronic devices and other boron-based nano-devices in the future. To our knowledge, experimental or theoretical investigations on the electronic transport properties of borophene are still lacking.

In this paper, we investigate the phonon-limited intrinsic electric resistivity of 2D metallic borophenes. Our simulations based on first principles yield intrinsic electric transport properties in these unique 2D metals. We calculate the electron-phonon contribution to the electric resistivity of borophenes as a function of temperatures and carrier densities, with phonon dispersions and electron-phonon couplings calculated within density-functional perturbation theory with very accurate Wannier interpolation. We demonstrate that the intrinsic resistivity of borophene is quite different from semimetallic graphene at low temperatures. The temperature-dependent $\rho_{e\text{-}ph}$ in borophenes is in good line with the Bloch-Grünesisen model. In addition, the resistivity can be highly tuned by charge carriers by applying substrate or external potential. This work not only provides a deep insight into the intrinsic transport properties of a representative 2D metal----borophene, but also provides a physical foundation for future applications of the emerging 2D materials.

**Theoretical Methodology**

The electron-phonon interaction matrix is computed using density functional perturbation theory as:

$$g_{mn,\nu}(k,k') = \frac{1}{\sqrt{2\omega_{k'-k\nu}}} \langle \psi_{mk'} | \partial_{k'-k\nu} V | \psi_{nk} \rangle \quad (1),$$

where $|\psi_{nk}\rangle$ is the electronic eigenstate of a Bloch state with the energy of $\varepsilon_{nk}$ (band index *n* and wave vector **k**), $\partial_{k'-k\nu}V$ is the change in the self-consistent potential induced by a phonon mode with the energy of $\hbar w^{\nu}_{k'-k}$ (branch index *v* and wavevector $k'-k$). Employing a first-principles interpolation method based on maximally localized Wannier functions as implemented in the Wannier90[34-37] and EPW package[38-43], we calculate the *e-ph* coupling elements $g_{mn,\nu}(k,k')$ on an ultradense grid spanning over the whole Brillouin zone, which is necessary to calculate the transport Eliashberg function accurately[44-45],

$$\alpha_{tr}^2 F(\omega) = \frac{1}{2} \sum_v \int_{BZ} \frac{dq}{\Omega_{BZ}} \omega_{qv} \lambda_{tr,qv} \delta(\omega - \omega_{qv}) \quad (2).$$

Here the mode-resolved coupling strength is defined as,

$$\lambda_{tr,qv} = \frac{1}{N(\varepsilon_F)\omega_{qv}} \sum_{nm} \int_{BZ} \frac{dk}{\Omega_{BZ}} |g_{mn,v}(k,k')|^2 \times \delta(\varepsilon_{nk} - \varepsilon_F)\delta(\varepsilon_{mk+q} - \varepsilon_F)(1 - \frac{v_{nk} \cdot v_{mk+q}}{|v_{nk}|^2}) \quad (3),$$

where $v_{nk} = \partial \varepsilon_{nk}/\partial k$ is the electron velocity. Finally, the electronic transport properties are calculated by the Boltzmann transport equation. Here, we adopt the simplest and most popular approximation---- the Ziman's resistivity formula [46]:

$$\rho(T) = \frac{4\pi m_e}{ne^2 k_B T} \times \int_0^\infty d\omega \hbar \omega * \alpha_{tr}^2 F(\omega) m(\omega, T)[1 + m(\omega, T)] \quad (4),$$

where $m(\omega,T)$ is the Bose-Einstein distribution; $e$ and $m_e$ are the charge and the mass of an electron, respectively. See the method part for more technical details.

## Results and Discussion

Figure 1 shows atomic structures of three types of boron sheets realized experimentally and the corresponding $\rho_{e-ph}$ as the function of temperature. The fully relaxed $\beta_{12}$ borophene is perfectly planar, in which boron atoms have different coordination numbers (CN), *i.e.*, CN = 4, 5, 6. The $\beta_{12}$ borophene has a rectangle primitive cell (Fig. 1a) with two lattice constants equal 2.93 Å and 5.07 Å. Different form the $\beta_{12}$ phase, the $\chi_3$ phase (with the lattice constant equals 4.45 Å) consists of only two types of boron atoms with CN = 4, 5, indicated in the rhombus in Fig. 1b. The third phase (namely, triangle borophene, shown in Fig. 1c) considered here is the close-packed one with a zigzag pattern (CN = 6, the lattice constants are 1.61 Å and 2.87 Å), leading to a lattice anisotropy. Among the three polymorphs, the $\beta_{12}$ and $\chi_3$ borophenes share very similar formation energy, while triangle borophene is energetically less stable (by 68 meV/atom).

We present in Figure 1(d-i) the phonon-limited resistivity $\rho_{e-ph}$ of the three borophene polymorphs in both linear and logarithmic scales. We see that, $\rho_{e-ph}$ of $\beta_{12}$ borophene is linear with $T^4$ at low temperatures ($T < 133$ K, illustrated in light blue region of panel d). In contrast, $\rho_{e-ph}$ is linear with $T$ when the temperature is large than 133 K, with a slope of 0.03 $\mu\Omega$*cm/K, as shown in the light red region. This observation reflects the 2D nature of electrons and phonons in the borophene. At room temperature, the intrinsic resistivity of $\beta_{12}$ borophene is 6.87 $\mu\Omega$*cm, in the same order of magnitude with graphene (1.0 $\mu\Omega$*cm)[5]. Another prominent feature of $\beta_{12}$ borophene is the emergence of transition at a very low temperature (~133 K), as indicated in Figure 2d.

As to $\chi_3$ and triangle borophenes, similar trends between the intrinsic resistivity $\rho_{e-ph}$ and temperature $T$ are observed (Figure 1e-f). In addition, the $\rho_{e-ph}$ of the two borophenes is larger than that of $\beta_{12}$ phases: the $\rho_{e-ph}$ at room temperature is 20.10 and 12.63 $\mu\Omega$*cm, respectively. We interpret the difference as that the $\rho_{e-ph}$ of different polymorphs of borophene are strongly dependent on their respective atomic structure, mainly attributed to difference in CNs and carrier densities. To have a direct comparison of the transition point, the crossover from $T^4$ to $T$ region locates at 96 K and 105 K for the $\chi_3$ and triangle borophene, respectively. The transition temperatures $\Theta_{BG}$ of three borophenes are all remarkably lower than the Debye temperature of borophene $\Theta_D$. The temperatures $\Theta_D$ are about 1700 K for all the three 2D boron sheets, which corresponds to the highest phonon energy ~1200 cm$^{-1}$. Therefore, we come to the first finding of this work: the temperature-dependent $\rho_{e-ph}$ of the borophenes are sensitive to the atomic structures, and agree well with the Bloch-Grünesisen model with a $\Theta_{BG}$ temperature of ~100 K.

In the following, we look into the contributions of different phonon branches to the intrinsic resistivity of borophene. The characteristic features of phonon-limited resistivity in the three borophenes are shown in Figure 2a-c. Here, we plot the total $\rho_{e\text{-}ph}$ of three polymorphs in linear scale together with the contributions from different phonon modes. For $\beta_{12}$ borophene, it is obvious that the contribution from the out-of-plane acoustic mode (#1: ZA, shown in Figure 2d) is the largest in the range of 0 K and 600 K, accounting for about 30% of the total $\rho_{e\text{-}ph}$. Transverse acoustic mode (#2: TA, shown in Figure 2g) and out-of-plane transverse optical mode (#4: TO, Figure 2j) with a frequency of 149 cm$^{-1}$ at $\Gamma$ point also play an important role in the phonon-mediated intrinsic resistivity, which are responsible for 20% and 15% of $\rho_{e\text{-}ph}$, respectively. The relative contributions of different modes are not sensitive to the variation of temperature (from 0 K to 600 K). Therefore, one can argue that low-temperature acoustic phonon modes are the main resources of total $\rho_{e\text{-}ph}$ at low temperatures. However, the contribution of TO phonon mode cannot be ignored even at low temperatures (*e.g.*, room temperature).

Different from $\beta_{12}$ borophene, ZA phonon modes (#1 in Figure 2e,f) of $\chi_3$ and triangle borophenes contribute a dominant part in the total $\rho_{e\text{-}ph}$ (~73 % and ~70 % for $\chi_3$ and triangle borophenes, respectively). It is clear that the optical phonon modes take up quite small fractions (~5% for both) of the total $\rho_{e\text{-}ph}$, which is illustrated in Figure 2b,c. It is reasonable since a higher excitation energy or much higher temperature is needed to excite the optical phonon modes. The observation that only low-energy phonons are the main contribution of total $\rho_{e\text{-}ph}$ is a direct evidence for the Bloch-Grüneisen behavior in the 2D metals.

For evaluating the intrinsic transport properties of 2D materials, deformation potential approximation proposed by Shockley and Bardeen[47] has been widely used, where only the LA phonons are considered to scatter carriers. The method has been applied to many 2D materials, *e.g.*, graphene and transition metal dichalcogenides.[4,48-49] However, our findings strongly suggest the failure of deformation potential theory for calculating electrical transport properties of borophene, since the contribution from ZA or TA mode of borophene is significantly important, similar to the behavior of 2D semimetallic silicene and stanene.[50]

To understand the electron-electron (*e-e*) interactions beyond local density approximation (LDA) in calculating transport properties, we compare the band structures calculated at the level of LDA and GW[51,52], respectively, shown in Figure S1 in Supporting Information (SI). Obviously, the band dispersion of the borophene from LDA is in good line with that incorporating quasi-particle interaction. On the other hand, Park *et al.* have studied the *e-e* interactions of graphene at the level of many-body perturbation theory and shown that only the contributions of high-energy optical phonons to the intrinsic resistivity is affected at high temperatures (~500 K) in GW approximation.[6] In metallic borophenes, low-energy acoustic phonon modes are dominant resources of the total $\rho_{e-ph}$ even at ~600 K, leading to a negligible influence in the intrinsic resistivity of *e-e* interactions even at the GW level. Consequently, our calculations based on LDA are accurate enough to obtain reasonable results in intrinsic resistivity.

Carrier density is an effective degree of freedom to manipulate electron-phonon interactions in two-dimensional materials. As mentioned earlier, the phonon-limited resistivity of graphene is highly tunable by adjusting carrier density, especially the transition temperature $\Theta_{BG}$ (ranging from 100 K to 1000 K)[5-6]. We notice that the charge doping effect from the silver substrates to $\beta_{12}$ and $\chi_3$ borophene is reported experimentally[17-21]. Charge transferred to the boron layer can greatly alter the Fermi level and the shape of Fermi surface. In addition, Zhang *et al.* have reported that gate voltage is able to control the energy-favored boron sheets, offering new insight to the relative stability of borophenes at different doping levels[15]. Therefore, an intriguing question arises as to what effects charge carriers can do to modulate the intrinsic electric resistivity of borophene.

We take $\beta_{12}$ borophene as an example to tune the electric transport performance under high carrier densities. Figure 3 summaries the intrinsic electric resistivity of $\beta_{12}$ borophene by adding additional electrons/holes ($n = \pm 2.0 \times 10^{14}$ cm$^{-2}$ to $\pm 3.3 \times 10^{14}$ cm$^{-2}$) to the systems. Here, we use "−" to represent electron doping and "+" to indicate hole doping. There is no imaginary phonon vibration for the carrier densities mentioned above, suggesting that the stability of 2D boron sheets can be preserved under ultrahigh carrier densities. It is somewhat surprising that $\rho_{e-ph}$ of three borophene polymorphs can be largely tuned by external charge carriers. Considering $n = +2.0 \times 10^{14}$ cm$^{-2}$, the $\rho_{e-ph}$ of $\beta_{12}$ borophene increased 2.7 times as much as that of pristine one (from 6.9 μΩ*cm to 18.3 μΩ*cm at room temperature). Furthermore, this value grows to 34.5 μΩ*cm (about 5 times over that of the pristine one) when the doping level increases to $n = +3.3 \times 10^{14}$ cm$^{-2}$.

In order to obtain a complete picture, we also provide the result of electron doping at the same doping levels. Our data in Figure 3a confirm the modulation in the $\rho_{e-ph}$ originating from the electron-phonon interactions, despite the changes in $\rho_{e-ph}$ are relatively lower (1.5 and 1.7 times larger than that of pristine $\beta_{12}$ borophene at $n = -2.0 \times 10^{14}$ cm$^{-2}$ and $-3.3 \times 10^{14}$ cm$^{-2}$, respectively).

To gain a quantitative analysis on the Bloch-Grüneisen behavior in $\beta_{12}$ borophenes at different densities $n$, we fit $\rho_{e-ph}$ at two distinct regimes (Figure 3b). The crossover between the two regimes exhibits a small variation, ranging from 75 K to 133 K for different doping densities.

We interpret the data as indicating that the size of Fermi surface in the borophene does not change appreciably when the ultrahigh carrier density is applied, displayed in Figure S2 (see SI). Despite a complex Fermi surface for $\beta_{12}$ borophene, it can be argued that only the electron/hole pocket at the center of the BZ is mainly responsible for the intrinsic transport properties. Surprisingly, our result reflects that borophene is significantly different from graphene in the carrier-tuned transport behavior. On one hand, the absolute value of $\rho_{e\text{-}ph}$ of borophene is highly sensitive to external carrier densities, while ultrahigh doping level in graphene leads to a quite small variation in $\rho_{e\text{-}ph}$. On the other hand, the Bloch-Grüneisen transition point $\Theta_{BG}$ is nearly fixed around 100 K with the variation of the carrier densities, while in graphene the Bloch-Grüneisen temperature changes dramatically, in a large range from 100 K to 1000 K with doping levels of $4 \times 10^{14}$ cm$^{-2}$. The fixed Bloch-Grüneisen transition temperature upon doping and atomic structure variations in different phases of borophene suggest that the intrinsic resistivity of borophene follows a rather stable universal scaling behavior with the temperature, for a large temperature range, which is desirable in many nanoelectronic applications.

The mobility ($\mu$) of borophene is another important quantity to understand the transport property of 2D metals. Here, we estimate the $\mu$ of $\beta_{12}$ borophene by the relationship: $\sigma = Ne\mu$, where $\sigma$ is the electric conductivity and $N$ is the carrier density. The carrier density for pristine $\beta_{12}$ borophene is estimated as $3.4 \times 10^{13}$ cm$^{-2}$, based on the proportion between the pocket area at the BZ center and the whole area of Fermi surface. At room temperature, the intrinsic mobility of $\beta_{12}$ borophene is estimated as 540 cm$^2$/(V*s), which is much smaller than that of graphene ~$10^5$ cm$^2$/(V*s)[53]. For the doped cases, we obtain the carrier densities by linearly dependence on the density of the state around Fermi level. The mobility decreases to ~100 and ~50 cm$^2$/(V*s) for $n = +2.0 \times 10^{14}$ cm$^{-2}$ and $+3.3 \times 10^{14}$ cm$^{-2}$, respectively. Notably, the value of $\mu$ is in inverse proportion with the assumed free carrier density and the method somewhat overestimates the carrier density. Nevertheless, the significant difference in the carrier-density modified mobility is robust and can be attributed to the modulations of electron-phonon interactions by external doping carriers.

More information comes from the carrier-mediated band structures and phonon dispersions, as shown in Figure 4. Hole doping lowers the Fermi energy by 0.30 eV (panel a) and 0.43 eV (panel b) at $n = +2.0 \times 10^{14}$ cm$^{-2}$ and $+3.3 \times 10^{14}$ cm$^{-2}$, respectively. In contrast to the simple linear performance of the Dirac bands around Fermi level in graphene, the electronic states around Fermi level vary more complicatedly, because of much complex state distributions (see Figure 4a). Importantly, the frequency of acoustic phonons (Figure 4d-f) with the lowest energy decreases along M-Y direction in the BZ (*e.g.*, from 132.6 cm$^{-1}$ to 63.4 cm$^{-1}$ at the middle of M-Y), indicating a phonon softening effect owning to hole doping. More clearly, the Eliashberg transport spectral functions (Figure 4, panel g-i) exhibit a strong enhancement, especially in the low-energy region (0-50 meV). This is a direct explanation for the large modulation in total $\rho_{e\text{-}ph}$ under different carrier densities.

The resistivity of borophene is not only very low (on the same order of magnitude as graphene for $\beta_{12}$ borophene) but also much easier to be tuned by carrier density while the linear dependence on *T* can be perfectly preserved for a considerably high doping level around room temperature. Considering it is difficult to tailor the carrier concentration in traditional bulk metals, 2D metallic borophene is an excellent platform for realizing highly carrier-dependent electronic transport devices. The above findings may yield new device applications for borophenes: the high carrier-density sensitivity can be utilized for an external-gate-regulating resistor or a memory resistor (or memristor, in which the resistivity varies with the historical accumulated carrier density)[54].

We note that the ultrahigh carrier level discussed above is reasonable and can be achieved for 2D systems in experiment, such as by an electrolytic gate or chemical absorption. In graphene, extremely high carrier densities (up to $4.0 \times 10^{14}$ cm$^{-2}$ for both electrons and holes) can be realized.[15] Indeed, one challenge at this stage to investigate the transport properties lie at that the borophene monolayer is hard to be exfoliated from the metallic substrates due to the relatively strong borophene-substrate interactions[14,17-18]. However, with the development of growth methods, we believe a freestanding (exfoliated) borophene will be realized soon in laboratory or experimental experts could find insulating substrates to grow such novel materials to validate the predictions reported above.

## Conclusion

To conclude, we employ *ab initio* calculations to investigate the phonon-limited intrinsic resistivity of borophenes with different polymorphs and carrier densities. Our study reveals the intrinsic resistivity of borophenes follows a linear relationship with $T^4$ at low temperatures. At high temperatures, $\rho_{e\text{-}ph}$ is linear with $T$. The resistivity is highly dependent on the polymorphs. Furthermore, the $\rho_{e\text{-}ph}$ of three borophene polymorphs can be greatly tuned by carrier densities. It is found that a Bloch-Grünesisen behavior with nearly fixed transition temperature is broadly satisfied at different temperatures and carrier densities. Furthermore, it is found that deformation potential approximation fails to describe electrical transport properties of borophene. Based on these findings, one could utilize different doping methods to control the resistivity of boron-based 2D metals, thus facilitating future applications in 2D electronic devices.

## Methods

We first use density-functional theory (DFT) and density-functional perturbation theory (DFPT) as implemented in the Quantum-ESPPRESSO package[38] with the local-density approximation (LDA)[39,40] to compute electronic and vibrational properties including *e-ph* coupling matrix elements.[41-45] Each borophene is separated from its periodic replicas by 15.0 Å to ensure that the effect of periodic boundary conditions is negligible. We use a kinetic energy cutoff of 80 Ry in expanding the valence electronic states in a plane wave basis and the core-valence interactions are taken into account by means of ultrasoft pseudopotentials[55]. Charge doping is modeled by adding extra electrons/holes and a neutralizing background. We use 30 × 20 × 1 (42 × 42 × 1 and 100 × 60 × 1) k-mesh in the full Brillouin integration for the charge density of $\beta_{12}$ ($\chi_3$ and triangle) borophene. The quantity $g_{mn,\nu}(k,k')$ was calculated first on a coarse grid of 6 × 4 × 1 (4 × 4 × 1 and 6 × 10 × 1) q-mesh in the Brillouin zone and then Wannier interpolated into an ultrafine grid of 300 × 200 × 1 (240 × 240 × 1 and 300 × 360 × 1) points for $\beta_{12}$ ($\chi_3$ and triangle) borophene.

### Acknowledgements

This work was supported by National Key Research and Development Program of China (Grant Nos. 2016YFA0300902 and 2015CB921001), National Natural Science Foundation of China (Grant Nos. 11474328 and 11290164), and "Strategic Priority Research Program (B)" of Chinese Academy of Sciences (Grant No. XDB07030100).

# References

1. Bloch, F., Zeitschrift für Physik **1930**, 59, 208.
2. Grüneisen, E., *Ann. Phys. (Leipzig)* **1933**,16, 530.
3. Fuhrer, M. S. *Physics* **2010**, 3, 106.
4. Hwang, E. H.; Das Sarma, S. *Phys. Rev. B* **2008**, 77, 115449.
5. Efetov, D. K.; Kim, P. *Phys. Rev. Lett.* **2010**, 105, 256805.
6. Park, C. H.; Bonini, N.; Sohier, T.; Samsonidze, G.; Kozinsky, B.; Calandra, M.; Mauri, F.; Marzari, N. *Nano Lett.* **2014**, 14, 1113−1119.
7. Kim, T. Y.; Park, C. H.; Marzari, N. *Nano Lett.* **2016**, 16 (4), 2439-43.
8. Novoselov, K.; Geim, A.; Morozov, S.; Jiang, D.; Katsnelson, M.; Grigorieva, I.; Dubonos, S.; Firsov, A. *Nature* **2005**, 438, 197.
9. Geim, A. K.; Novoselov, K. S. *Nat. Mater.* **2007**, 6, 183.
10. Zhang, Z.; Penev, E. S.: Yakobson, B. I. *Nat. Chem.* **2016**, 8, 525.
11. Wu, X.; Dai, J.; Zhao, Y.; Zhuo, Z.; Yang, J.; Zeng, X. C. *ACS Nano* **2012**, 6, 7443.
12. Penev, E. S.; Bhowmick, S.; Sadrzadeh, A.; Yakobson, B. I. *Nano Lett.* **2012**, 12 (5), 2441-2445.
13. Liu, Y.; Penev, E. S.; Yakobson, B. I. *Angew. Chem., Int. Ed.* **2013**, 52, 3156.
14. Zhang, Z.; Yang, Y.; Gao, G.; Yakobson, B. I. *Angew. Chem. Int. Ed.* **2015**, 54, 13022.
15. Zhang, Z.; Shirodkar, S. N.; Yang, Y.; Yakobson, B. I. *Angew. Chem. Int. Ed.* **2017** (DOI: 10.1002/anie.201705459).
16. Penev, E. S.; Kutana, A.; Yakobson, B. I. *Nano Lett.* **2016**, 16, 2522.
17. Mannix, A. J.; Zhou, X.-F.; Kiraly, B.; Wood, J. D.; Alducin, D.; Myers, B. D.; Liu, X.; Fisher, B. L.; Santiago, U.; Guest, J. R. *Science* **2015**, 350, 1513.
18. Feng, B.; Zhang, J.; Zhong, Q.; Li, W.; Li, S.; Li, H.; Cheng, P.; Meng, S.; Chen, L.; Wu, K. *Nat. Chem.* **2016**, 8 (6), 563-568.
19. Feng, B.; Zhang, J.; Liu, R.-Y.; Iimori, T.; Lian, C.; Li, H.; Chen, L.; Wu, K.; Meng, S.; Komori, F.; Matsuda, I. *Phys. Rev. B* **2016**, 94 (4).
20. Zhang, Z.; Mannix, A. J.; Hu, Z.; Kiraly, B.; Guisinger, N. P.; Hersam, M. C.; Yakobson, B. I., *Nano Lett.* 2016, 16 (10), 6622-6627.
21. Feng, B.; Sugino, O.; Liu, R.-Y.; Zhang, J.; Yukawa, R.; Kawamura, M.; Iimori, T.; Kim, H.; Hasegawa, Y.; Li, H.; Chen, L.; Wu, K.; Kumigashira, H.; Komori, F.; Chiang, T. C.; Meng, S.; Matsuda, I. *Phys. Rev. Lett.* **2017**, 118 (9), 096401.
22. Feng, B.; Zhang, J.; Ito S.; Arita M.; Cheng C.; Chen, L.; Wu, K.; Komori, F.; Sugino, O.; Kawamura, M.; Okuda T.; Meng, S.; Matsuda, I. *Adv. Mat.* **2017** (DOI: 10.1002/adma.201704025).
23. Gao, M.; Li, Q. Z.; Yan, X. W.; Wang, J. *Phys. Rev. B* **2017**, 95 024505.
24. Zhao, Y.; Zeng, S.; Ni, J. *Appl. Phys. Lett.* **2016**, 108, 242601.
25. Cheng, C.; Sun, J.-T.; Liu, H.; Fu, H.-X.; Zhang, J.; Chen, X. R.; Meng, S. *2D Materials* **2017**, 4 (2), 025032.
26. Padilha, J. E.; Miwa, R. H.; Fazzio, A. *Phys. Chem. Chem. Phys.* **2016**, 18 25491.

27. Lherbier, A.; Botello-Méndez, A. R.; Charlier, J.-C. *2D Mater.* **2016**, 3, 045006.
28. Zhou, X. F.; Dong, X.; Oganov, A. R.; Zhu, Q.; Tian, Y.; Wang, H. T. *Phys. Rev. Lett.* **2014**, 112, 085502.
29. Sun, X.; Liu, X.; Yin, J.; Yu, J.; Li, Y.; Hang, Y.; Zhou, X.; Yu, M.; Li, J.; Tai, G. *Adv. Funct. Mater.* **2017**, 27 (19).
30. Jiang, H. R.; Lu, Z.; Wu, M. C.; Ciucci, F.; Zhao, T. S. *Nano Energy* **2016**, 23, 97–104.
31. Cui, Z. H.; Jimenez-Izal, E.; Alexandrova, A. N. *J. Phy. Chem. Lett.* **2017**, 8 (6), 1224-1228.
32. Zhang, Z.; Yang, Y.; Penev, E. S.; Yakobson, B. I. *Adv. Funct. Mater.* **2017**, 27 (9), 1605059.
33. Huang, Y.; Shirodkar, S. N., and Yakobson, B. I. *J. Am. Chem. Soc.* DOI**:** 10.1021/jacs.7b10329
34. Marzari, N.; Vanderbilt, D. *Phys. Rev. B* **1997**, 56, 12847.
35. Souza, I.; Marzari, N.; Vanderbilt, D. *Phys. Rev. B* **2001**, 65, 035109.
36. Marzari, N.; Mostofi, A. A.; Yates, J. R.; Souza, I.; Vanderbilt, D. *Rev. Mod. Phys.* **2012**, 84, 1419.
37. Mostofi, A. A.; Yates, J. R.; Lee, Y.-S.; Souza, I.; Vanderbilt, D.; Marzari, N. *Comput. Phys. Commun.* **2008**, 178, 685.
38. Giannozzi, P.; et al. *J. Phys.: Condens. Matter.* **2009**, 21, 395502.
39. Ceperley, D. M.; Alder, B. J. *Phys. Rev. Lett*. **1980**, 45, 566.
40. Perdew, J. P.; Zunger, A. *Phys. Rev. B* **1981**, 23, 5048.
41. Giustino, F.; Cohen, M. L.; Louie, S. G. *Phys. Rev. B* **2007**, 76, 165108.
42. Giustino, F. *Rev. Mod. Phys*. **2017**, 89, 1.
43. Grimvall G., The Electron–Phonon Interaction in Metals, North-Holland Publishing Company, **1981**.
44. Noffsinger, J.; Giustino, F.; Malone, B. D.; Park, C.H.; Louie, S. G.; Cohen, M. L. *Comput. Phys. Commun*. **2010**, 181, 2140.
45. Poncé, S.; Margine, E. R.; Verdi, C.; Giustino, F. *Comput. Phys. Commun*. **2016**, 209, 116-133.
46. J. Ziman, Electrons and Phonons, *Oxford University Press*, **1960**.
47. Bardeen, J.; Shockley, W., *Phys. Rev.* **1950**, 80 (1), 72.
48. Kaasbjerg, K.; Thygesen, K. S.; Jacobsen, K. W., Phys. Rev. B 2012, 85 (16), 165440.
49. Cai, Y.; Zhang, G.; Zhang, Y.-W., *J. Am. Chem. Soc.* **2014**, 136 (17), 6269-6275.
50. Nakamura, Y.; Zhao, T.; Xi, J.; Shi, W.; Wang, D.; Shuai, Z., *Adv. Electron. Mater. 2017* (DOI: 10.1002/aelm.201700143).
51. Lazzeri, M.; Attaccalite, C.; Wirtz, L.; Mauri, F. *Phys. Rev. B* **2008**, 78, 081406.
52. Grüneis, A.; Serrano, J.; Bosak, A.; Lazzeri, M.; Molodtsov, S. L.; Wirtz, L.; Attaccalite, C.; Krisch, M.; Rubio, A.; Mauri, F.; Pichler, T. *Phys. Rev. B* **2009**, 80, 085423.
53. Morozov, S.; Novoselov, K.; Katsnelson, M.; Schedin, F.; Elias, D.; Jaszczak, J. A.; Geim, A., *Phys. Rev. Lett*. **2008**, 100 (1), 016602.
54. Strukov, D. B.; Snider, G. S.; Stewart, D. R.; Williams, R. S. *Nature* **2008**, 453 (7191), 80-83.
55. Vanderbilt, D. Phys. Rev. B **1990**, 41 (11), 7892.

## Figures and Captions

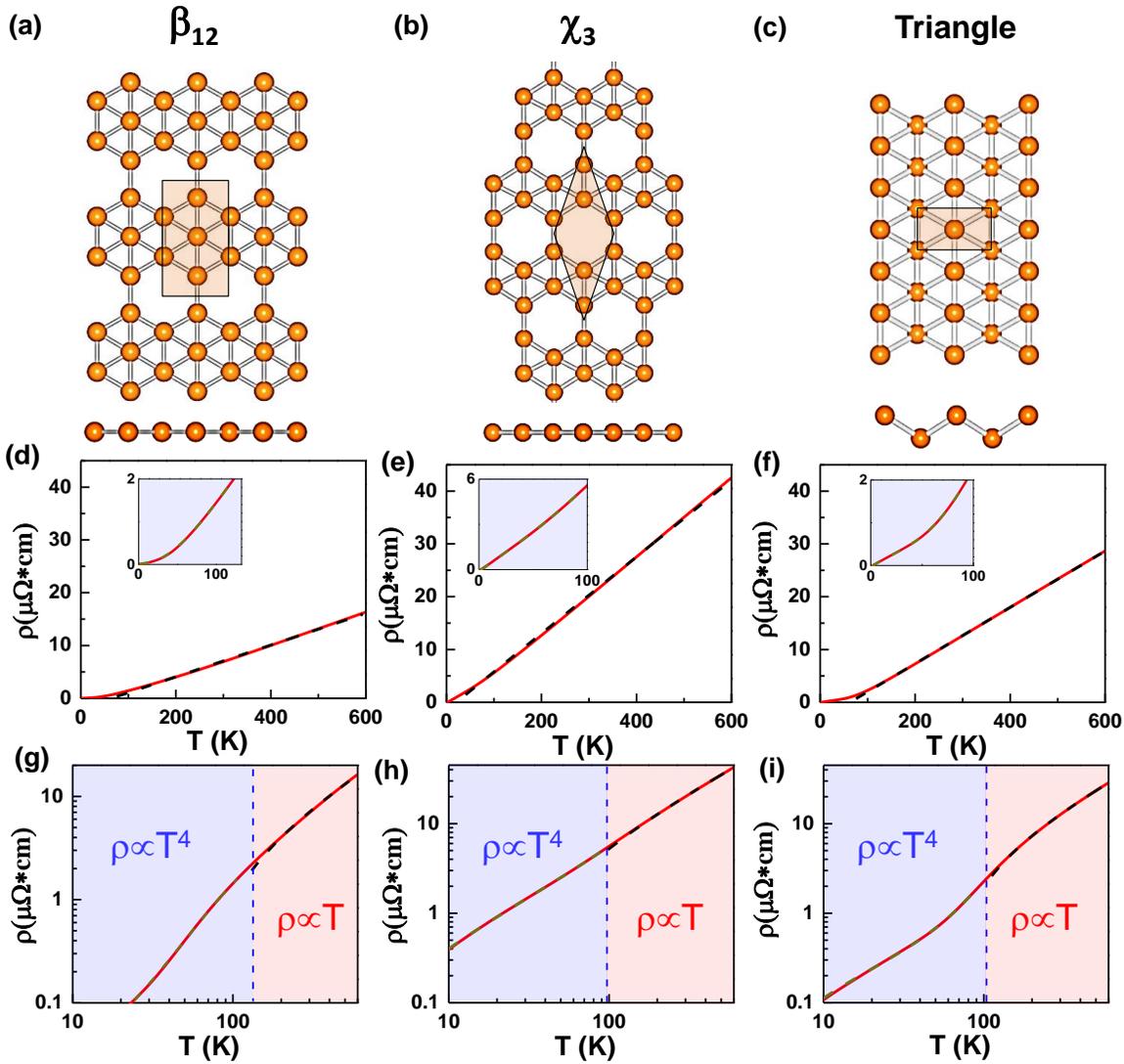

**Figure 1.** (a-c) Atomic structures of selected two-dimensional borophenes with the unit cells (top and side views). (a) $\beta_{12}$ borophene, (b) $\chi_3$ borophene, (c) triangle borophene. (d-f) Electrical resistivity of the three borophene in linear scale. The insets (light blue regimes) in (d-f) show the resistivity in smaller ranges at low temperatures. (g-i) Electrical resistivity of the three borophenes in logarithmic scale. The vertical dashed blue lines in (d-i) indicate the crossover between two regions. Light red regions show $\rho \propto T$, while light blue regions indicate $\rho \propto T^4$.

Solid red lines are calculated from first principle method and dashed black lines are fitted by linear functions. The dashed dark yellow lines are fitted by polynomial functions with an order of 4.

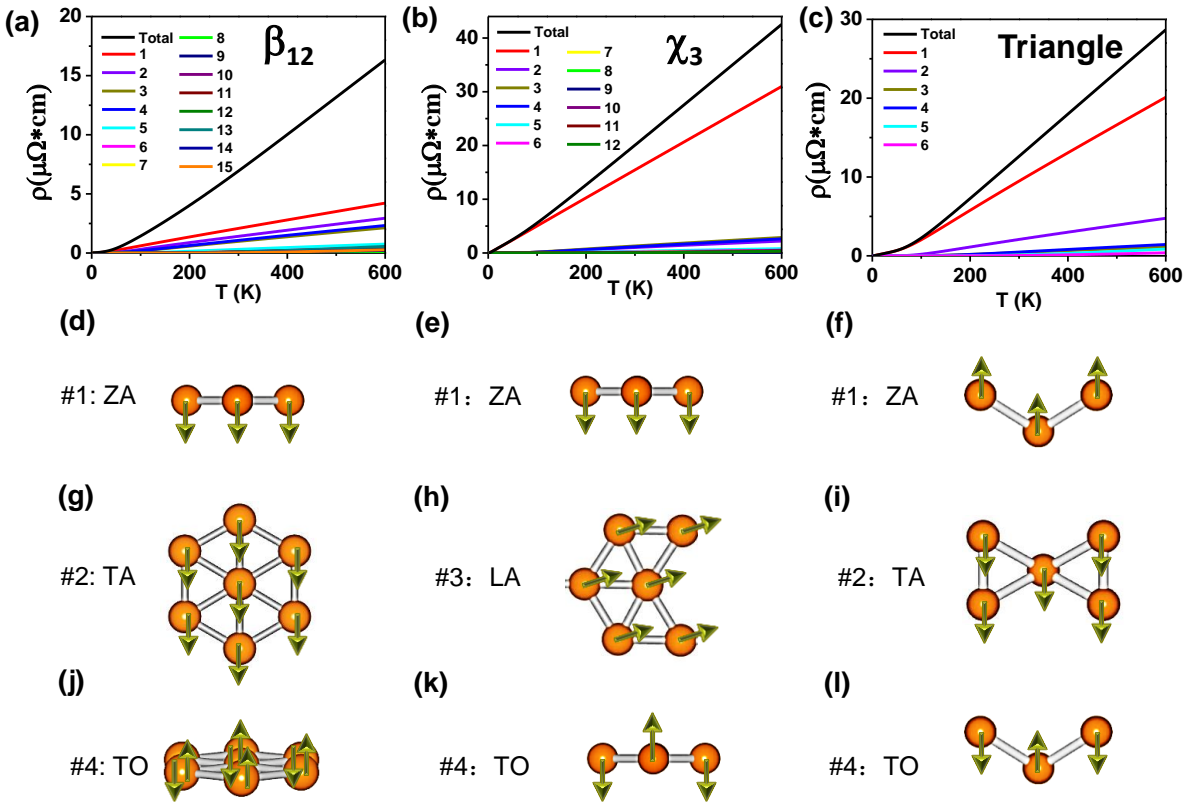

**Figure 2.** (a-c) Electrical resistivity of the three borophene polymorphs and the partial resistivity arising from each phonon branch in linear scale. (d-l) Atomic displacements of dominant phonon branches attributing to the phonon-limited resistivity. The sequences (#1, #2, *etc*.) are based on the relative energies of phonon modes. ZA, TA, LA and TO labeled in (d-l) indicate out-of-plane accoustic mode, transverse accoustic mode, longtudinal accoustic mode and transverse optical mode, respectively.

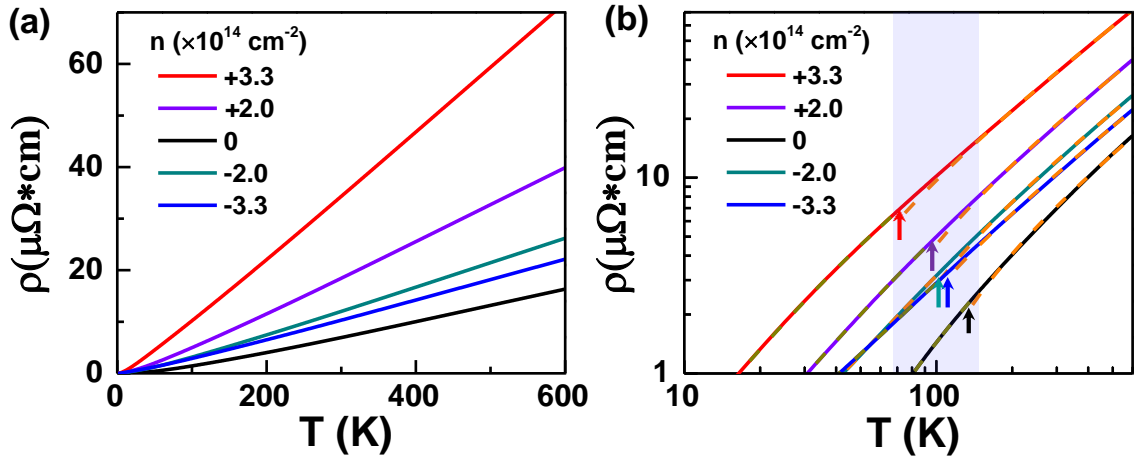

**Figure 3.** Electrical resistivity of β$_{12}$ borophene with different charge carrier densities ($n = \pm 2.0 \times 10^{14}$ cm$^{-2}$ and $n = \pm 3.3 \times 10^{14}$ cm$^{-2}$), (a) in linear scale and (b) in logarithmic scale. Solid lines are calculated from first-principles method and dashed orange lines are fitted by linear functions. The dashed dark yellow lines are fitted by polynomial functions with an order of 4. Arrows in different colors are given to illustrate the transition points for two different transport behaviors. We use "−" to represent electron doping and "+" to indicate hole doping.

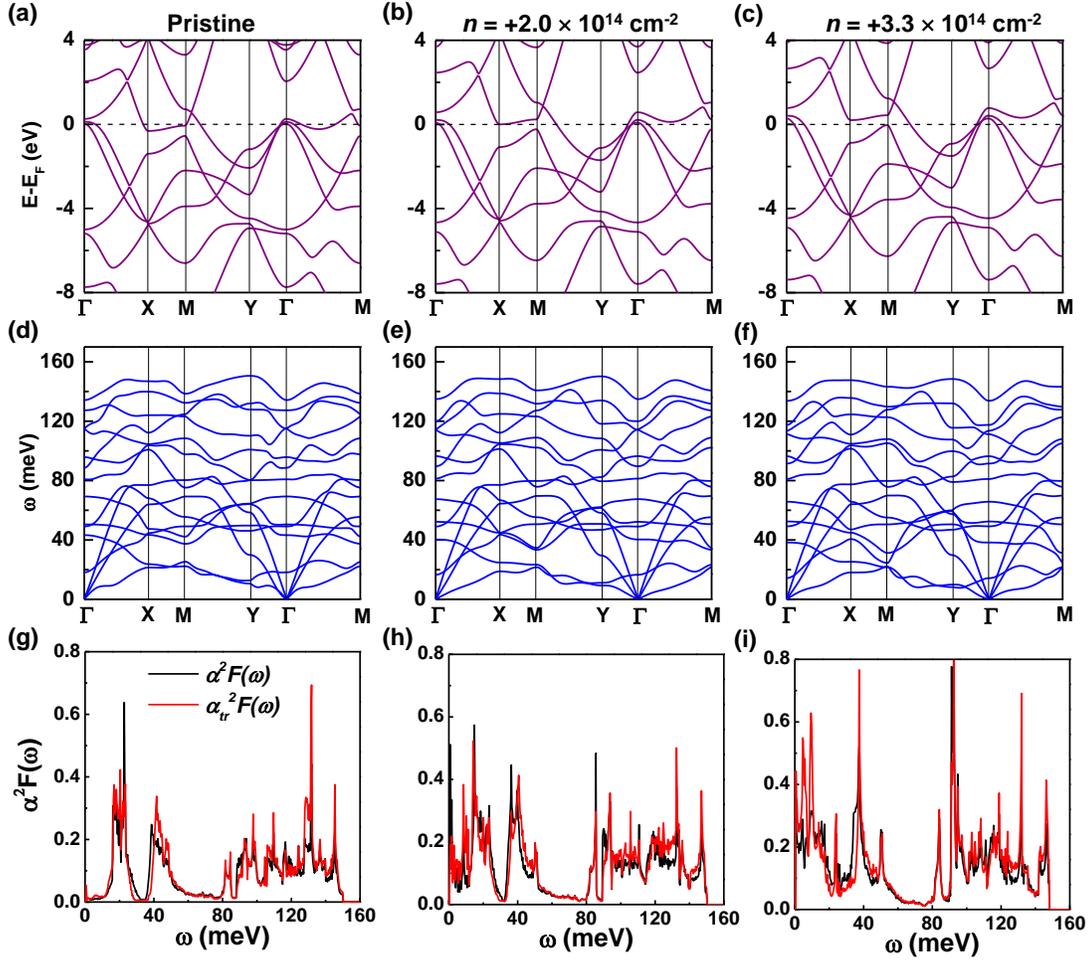

**Figure 4**. (a-c) Energy band structures of β$_{12}$ borophene at different hole densities (from $n = 0$ to $n = +3.3 \times 10^{14}$ cm$^{-2}$). Fermi levels are set to zero. (d-f) Phonon dispersions of β$_{12}$ borophene at different hole densities. (g-i) Corresponding Eliashberg spectral function $\alpha^2 F(\omega)$ (black lines), along with the Eliashberg transport spectral functions $\alpha_{tr}^2 F(\omega)$ (red lines).

**Table of Contents Graphic**

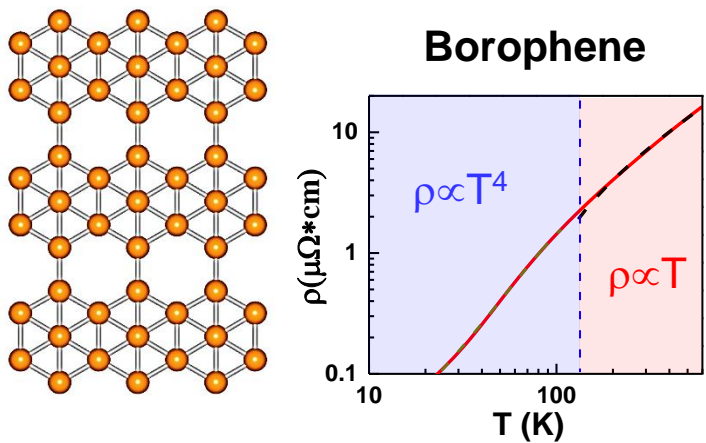

# Supporting Information for "Highly-Tunable Intrinsic Electrical Resistivity of Two-Dimensional Metallic Borophene"

Jin Zhang[1, 4, †], Jia Zhang[1, 4, †], Cai Cheng[1], Johannes Lischner[3], Feliciano Giustino[2,*], and Sheng Meng[1,4,5,*]

**S1, Band structures of $\beta_{12}$ borophene at LDA and GW0 functional**

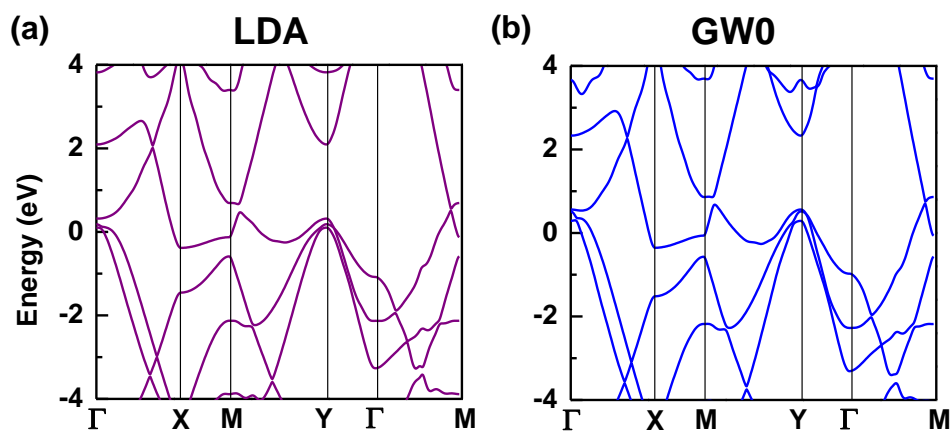

**Figure S1.** Band structures of $\beta_{12}$ borophene calculated using LDA (a) and GW0 functional (b), respectively. It is apparent that the shape and relative band energies of bands at LDA level are in good consistence with GW0 calculations.

## S2, Energy contours of $\beta_{12}$ borophene at different energies

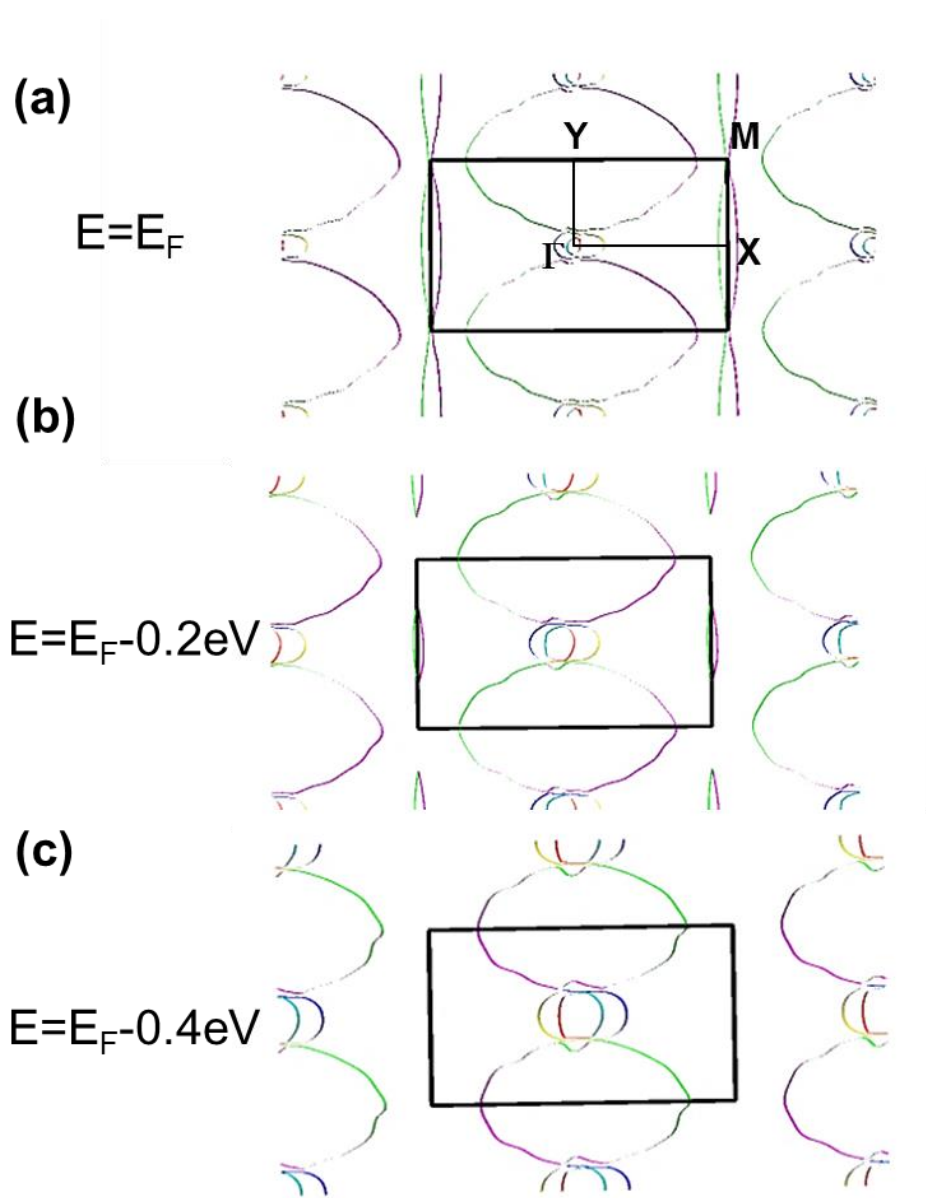

**Figure S2.** Energy contours of $\beta_{12}$ borophene at different energies ($E_F$, $E_F$-0.2 eV and $E_F$-0.4 eV). We can see the doing levels do not change the Fermi surfaces greatly, which explains the small variation of the temperature $\Theta$.